\documentclass[sigconf, authorversion]{acmart}

\usepackage{csvsimple}
\usepackage{threeparttable}

\usepackage{amsmath}
\usepackage{amsfonts}
\usepackage{multirow}

\usepackage[ruled,vlined,linesnumbered]{algorithm2e}
\DontPrintSemicolon
\SetArgSty{}
\SetKw{Continue}{continue}
\SetKw{Break}{break}
\SetKw{Print}{print}

\usepackage{listings}

\usepackage{pifont}

\usepackage[frozencache]{minted}

\copyrightyear{2022}
\acmYear{2022}
\setcopyright{acmcopyright}\acmConference[ICSE-NIER'22]{New Ideas and Emerging Results }{May 21--29, 2022}{Pittsburgh, PA, USA}
\acmBooktitle{New Ideas and Emerging Results (ICSE-NIER'22), May 21--29, 2022, Pittsburgh, PA, USA}
\acmPrice{15.00}
\acmDOI{10.1145/3510455.3512786}
\acmISBN{978-1-4503-9224-2/22/05}

\begin{document}

\setlength{\belowdisplayskip}{3pt}
\setlength{\abovedisplayskip}{3pt}
\title{Runtime Prevention of Deserialization Attacks}

%
\author{Fran\c{c}ois Gauthier}
\email{francois.gauthier@oracle.com}
\affiliation{%
  \institution{Oracle Labs}
  \city{Brisbane}
  \state{Queensland}
  \country{Australia}
}

\author{Sora Bae}
\email{sora.bae@oracle.com}
\affiliation{%
	\institution{Oracle Labs}
	\city{Brisbane}
  	\state{Queensland}
	\country{Australia}
}


\begin{abstract}
Untrusted deserialization exploits, where a serialised object graph is used
to achieve denial-of-service or arbitrary code execution, have become so prominent
that they were introduced in the 2017 OWASP Top 10. In this paper, we present a 
novel and lightweight approach for runtime 
prevention of deserialization attacks using Markov chains. The intuition
behind our work is that the \emph{features} and \emph{ordering} of classes in malicious 
object graphs make them distinguishable from benign ones. Preliminary results indeed
show that our approach achieves an F1-score of 0.94 on a dataset of 264 serialised 
payloads, collected from an industrial Java EE application server and a repository of
deserialization exploits.

\end{abstract}

%



\maketitle

\begin{figure}
	\includegraphics[width=\columnwidth]{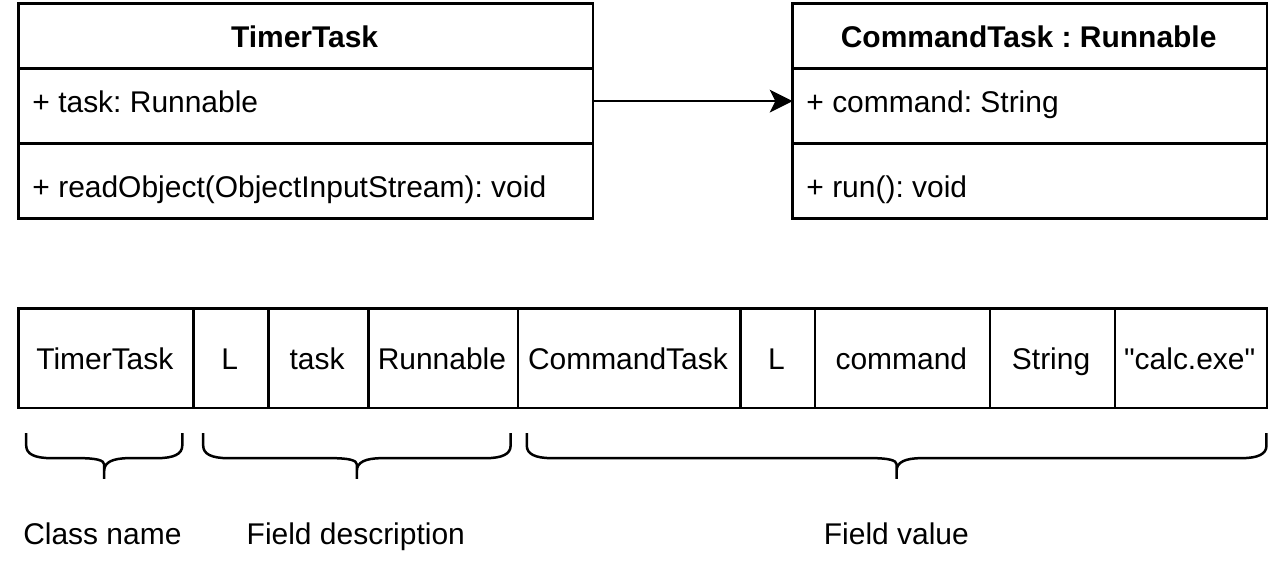}
	\caption{A simple class diagram and its corresponding byte stream}
	\label{fig:class_diagram}
\end{figure}

\section{Introduction}
In programming languages, serialization is the process of converting an in-memory object or data 
structure into a persistent format; deserialization works the opposite way. An attacker accessing
the serialized form of an object can thus influence the object that will be created upon
deserialization. In recent years, security researchers discovered various ways of
abusing deserialization to achieve denial-of-service or arbitrary code execution in various
languages like Java,
C\#, PHP, Python, and Ruby using various serialization formats like
binary, XML, JSON, and YAML~\cite{ysoserial, ysoserial2, marshall, Deser.NET}. Deserialization
issues are now so prominent that they are now included in the OWASP Top 10 Web Application 
Security Risks list~\cite{owasp.top10}. In this paper, we focus on detecting attacks against 
native Java deserialization that uses byte streams as the serialization format.
To help combat the threat posed by native Java deserialization vulnerabilities, deserialization 
filters were introduced in Java 9 and back-ported to Java 6, 7, and 8~\cite{jep290}. Upon
deserialization of, the filter is invoked after resolving the class from the byte
stream and before creating an object of that class in memory, giving the filter an opportunity
to inspect the class and stop the deserialization process altogether if a forbidden class is
detected. However, the onus of developing the filters is on developers and the manual effort
involved is not trivial. In this paper, we propose an approach to automatically build 
probabilistic models from \emph{benign} and \emph{malicious} byte streams to detect malicious 
deserialization at runtime.

\section{Background on Java deserialization}

In Java, any object from a class that directly or indirectly (i.e. through inheritance) implements 
the \mintinline{java}{Serializable} interface can be serialized and deserialized using Java 
native serialization. During the serialization process, starting from the root object, references 
to other objects (e.g. through class fields) are resolved and serialized deterministically in a 
recursive manner until the entire object graph has been converted to a byte stream. During
deserialization, the serialized object graph is read sequentially from the byte stream, one object
at the time. When an object is deserialized, the first information that is extracted from the stream 
is its class name, at which point deserialization filters are invoked to give the 
application an opportunity to introspect the class and interrupt deserialization if desired.
Figure~\ref{fig:class_diagram} shows an example of a class diagram (top) with a simplified
representation of its corresponding byte stream (bottom). 
Assuming that instances of
those classes have been constructed, serializing the corresponding object graph would
result in the byte stream at the bottom of Figure~\ref{fig:class_diagram}. The first part of the 
byte stream contains the name of the 
class of the first object to be deserialized, followed by field descriptions. The last part of
the stream contains the field values, which can be objects themselves. If this stream
was deserialized, the deserialization filter would receive the \mintinline{java}{TimerTask} class
first, followed by the \mintinline{java}{CommandTask} class.
Detailing the exact mechanisms that an attacker can use to exploit Java deserialization is beyond 
the scope of this paper, but the interested reader can refer to~\cite{TBS, ysoserial, ysoserial2} 
for more information. The key takeaway is that object graphs are serialized, deserialized, and 
filtered sequentially; a property we leverage to abstract them as Markov chains.

\section{Background on Markov Chains}
\label{sec:markov}

A Markov chain represents a system that has a finite number of states:
$S = \{s_1,s_2,\ldots,s_n\}$
and that transitions between states with some probability $p$ at each step $t$. 
The probability of the system starting in a state $s_i \in S$ is captured by its initial state 
probability vector: $p_{init} = (p_1, p_2, \ldots, p_n)$,
where each probability $p_i$ corresponds to the probability of the chain starting in state $s_i$
and where the probabilities in $p_{init}$ sum to one.
In Markov chains, the probability of transitioning from a state $s_i$ to another state $s_j$ depends 
on $s_i$ only, and is captured by a transition probability matrix, 
where rows correspond to the state at step $t$, columns correspond to the state at step $t+1$, and each 
row sums to one:
\begin{equation*}
p_{tr} = 
\begin{pmatrix}
p_{11} & p_{12} & \ldots & p_{1n} \\
\vdots & \vdots & \ddots & \vdots \\
p_{n1} & p_{n2} & \ldots & p_{nn} \\
\end{pmatrix}
\end{equation*}
Given a Markov chain
and a sequence of states $(x_1,x_2,\ldots,x_n)$, one can calculate the probability that the chain 
\emph{generated} the sequence with a simple product of probabilities:
\begin{equation}
P((x_1,x_2,\ldots,x_n)) = p_{init}(x_1) \cdot \prod_{i=2}^{n} p_{tr}(x_{i-1}x_{i})
\label{eq:probability}
\end{equation}

\section{Modelling Java Deserialization with Markov Chains}
In this section, we explain how we model deserialization as a Markov 
chain. Our choice of Markov chains over other
learning approaches was motivated by two main factors: 1) previous failed experiments with a sequence-agnostic classifier, and 2) the need to make predictions based on small datasets. We indeed 
experimented with na\"ive Bayes classifiers first and only achieved marginal improvement over random 
classification. Then, we chose Markov chains (MC) over more complex sequence-based learning approaches
that require large datasets like RNN, and LSTM because in our setup, deserialization is a relatively 
uncommon operation, leading to a small dataset. In \autoref{sec:results}, we show how MC generated from 
Bayesian inference, which accounts for the inherent uncertainty of small datasets, lead to significantly 
more precise predictions than MC that are derived directly from empirical data.

\subsection{Abstracting Java classes as states in a Markov chain}
Classes in a stream and their various features determine the code that is executed during
deserialization. For example, in Figure~\ref{fig:class_diagram},
because the \mintinline{java}{CommandTask} class implements the \mintinline{java}{Runnable}
interface, the \mintinline{java}{CommandTask} constructor will be invoked to create
a \mintinline{java}{Runnable} object for the \mintinline{java}{task} field. Also, because
the \mintinline{java}{TimerTask} class overrides the \mintinline[breaklines]{java}{readObject} method,
it can alter default deserialization, which is a feature that is often exploited in deserialization attacks.

To exploit a deserialization vulnerability, attackers will seek to \emph{craft} a byte stream consisting of 
specific classes with specific features in a specific order that will typically lead to denial-of-service
or arbitrary code execution. The key insight
here is that the class's intended use is largely irrelevant to attackers, who are instead solely 
interested in the very specific features that will lead to successful exploitation. For this reason,
we studied all the deserialization exploits in \texttt{ysoserial} \cite{ysoserial.git} and manually identified some of 
the features that make them \emph{useful} for exploitation purposes. The non-exhaustive list of 
features we identified are listed in Table~\ref{tab:features}. 

Given the set of features in Table~\ref{tab:features}, we can abstract each class as a Boolean 
feature vector. Given $n$ Boolean features, the number
of possible feature vectors is finite and equal to $2^n$. In our setup, the set of states in the 
Markov chain is thus the set of possible feature vectors. Because the maximum number of states grows 
exponentially with the number of features, in practice, we use the set of feature vectors that are 
observed in the training set, which cardinality tends to be much smaller than $2^n$. To account 
for new feature vectors in the testing set, we create a generic state to which all unobserved 
states map to.

\begin{table*}
	\caption{Common class features used in deserialization exploits}
	\centering
	\begin{tabular}{cll}
		\toprule
		Id & \multicolumn{1}{c}{Feature} & \multicolumn{1}{c}{Description} \\
		\midrule
		\multirow{4}{*}{1} & \multirow{4}{*}{Uses reflection} & True if the class calls any of the following from \mintinline{java}{java.lang.reflect}: \\
		& & ~~~~- \mintinline{java}{Constructor.newInstance()} \\
		& & ~~~~- \mintinline{java}{Field.set()} \\
		& & ~~~~- \mintinline{java}{Method.invoke()}\\
		\midrule
		2 & Overrides readObject & True if the class overrides the method \mintinline{java}{Object readObject(ObjectInputStream ois)}\\
		\midrule
		3 & Overrides hashCode & True if the class overrides the \mintinline{java}{int hashCode() method.}\\
		\midrule
		\multirow{4}{*}{4} & \multirow{4}{*}{Has generic field} & True if the class has a field of any of the following type:\\
		& & ~~~~- \mintinline{java}{java.lang.Object} \\
		& & ~~~~- \mintinline{java}{java.lang.Comparable} \\
		& & ~~~~- \mintinline{java}{java.util.Comparator} \\
		\midrule
		5 & Implements Map & True if the class implements the \mintinline{java}{java.util.Map} interface.\\
		\midrule
		6 & Implements Comparator & True if the class implements the \mintinline{java}{java.util.Comparator} interface.\\
		\midrule
		\multirow{4}{*}{7} & \multirow{4}{*}{Calls hashCode} & True if the class calls any of the following methods:\\
		& & ~~~~- \mintinline{java}{int java.util.Objects.hash(Object... values)}\\
		& & ~~~~- \mintinline{java}{int java.util.Objects.hashCode(Object o)}\\
		& & ~~~~- \mintinline{java}{*.hashCode()}\\
		\midrule
		\multirow{3}{*}{8} & \multirow{3}{*}{Calls compare} & True if the class calls any of the following methods:\\
		& & ~~~~- \mintinline{java}{*.compare()} \\
		& & ~~~~- \mintinline{java}{*.compareTo(...)} \\
		\bottomrule
	\end{tabular}
	\label{tab:features}
\end{table*}

\subsection{Estimating probabilities from data}
\label{sec:inference}
We estimate the various probabilities in our Markov chain directly from dynamic observations. 
Specifically, given a set of byte streams, we deserialize them, extract the resulting sequences 
of classes, and abstract all classes to feature vectors. This results in a set of concrete Markov 
chain instances (i.e. sequences of states) from which we can estimate the initial and state transition
probabilities. The most straightforward approach is to directly use empirically observed frequencies 
as probabilities. This works well in contexts where the sample size is large. When the sample size is small, however, statistical inference methods are generally 
preferable. 
In this work, we use Bayesian inference to estimate the probabilities of a Markov chain 
where empirical observations are used to guide the inference process. Bayesian inference models 
the variables to infer as random variables 
issued from specific probability distributions. Then, through a guided random process (e.g. 
Markov Chain Monte Carlo), it infers the parameters of those probability distributions that 
maximise the likelihood of the empirically observed values. Consider, for example, the transition probability matrix of a Markov chain. Our goal is to 
estimate the transition probabilities that best explain the observed sequences of states. 
A typical way of modelling such a matrix is to represent each row as the 
outcome of a Dirichlet distribution, which is parameterised with a vector of concentration 
parameters $(\alpha_1,\ldots,\alpha_K)$ where $\alpha_i > 0$ and produces as output 
a vector of $K$ real numbers that sum to one:
\begin{equation*}
(x_1,\ldots,x_K), \text{ where } x_i \in [0,1], \text{and } \sum_{i=1}^{K}x_i = 1
\end{equation*}
The concentration vector is used to initialise the distribution and captures our \emph{prior} 
knowledge about transition probabilities. In our setup, all the concentration parameters are 
set to one, to represent that we have no prior knowledge.
By repeatedly adjusting the concentration parameters, sampling from the per-row Dirichlet 
distributions, and evaluating the likelihood of the resulting matrix against our observations,
Bayesian inference eventually converges to a set of \emph{likely} transition matrices.
It is important to note that through its inference process, Bayesian inference actually 
generates multiple distributions for each row, where the more recent ones are expected to 
more precisely capture the \emph{real} probabilities. In cases where the observations are few, 
the inference might not converge to a single solution, but rather to a set of plausible 
solutions. We later use metrics like standard deviation over the set of generated 
solutions to estimate the \emph{confidence} in our predictions.

\section{Runtime Prevention of Deserialization Attacks}

We now present our approach to infer Markov chains from benign and malicious deserialization examples 
and predict if a given byte stream is malicious. In our setup, the benign and malicious examples come
from the application under test and from the \texttt{ysoserial} dataset~\cite{ysoserial,ysoserial.git}
respectively. To collect the classes and their associated features, we implemented
a custom deserialization filter that uses ASM~\cite{bruneton2002asm} to dynamically 
extract features from deserialised classes. To enable the classification of byte streams as 
benign or malicious, we create two Markov chains during the inference phase: one from benign 
examples, and one from malicious examples. Once the inference phase is complete, we use another 
custom deserialization filter to detect and prevent deserialization attacks based on the inferred
Markov chains. A simplified filter is illustrated in \autoref{alg:predict}. It takes as input the benign 
($\mathcal{B}$) and malicious ($\mathcal{M}$) Markov chains as well as two threshold parameters 
$t$ and $l$. Once deserialization starts, the Java runtime invokes the filter every time a new 
class is read from the serialized stream and passes it the class and a Boolean flag indicating 
whether the end of the stream has been reached. In practice, we must derive the $end$ flag from
other filter inputs, but we omit these details for clarity. The filter then uses 
ASM~\cite{bruneton2002asm} to abstract the class as a feature vector and appends it to the current 
sequence of classes (lines 3-4). It then computes the mean probability that the sequence has 
been \emph{generated} by the benign or malicious Markov chains (lines 5-6). Then, it computes 
confidence intervals of $t$ standard deviations around the means and checks if the intervals are
disjoint (line 7). If the end of the stream has been reached and the intervals are disjoint, the
highest mean probability determines the outcome (line 9). If the end has been reached but the 
intervals are not disjoint, we do not have enough confidence in the results to reach a decision and conservatively 
reject the stream (line 11). If the intervals are disjoint, and at least $l$ classes have been 
read from the stream, deserialization is aborted early if the stream is malicious (line 13). 
Otherwise, the filter postpones the decision and lets deserialization proceed to the next class 
in the stream by returning \texttt{undecided}.


\section{Preliminary Results}
\label{sec:results}

\begin{algorithm}
	\SetAlgoLined
	\SetKwProg{Def}{def}{:}{}
	\SetKwProg{Fn}{Function}{:}{end}
	\KwIn{$\mathcal{B}$, $\mathcal{M}$, $t$, $l$}
	\KwOut{status $\in$ \{\texttt{accepted}, \texttt{rejected}, \texttt{undecided}\}}
	$seq \leftarrow$ new List()\;
	\Fn{MarkovFilter(class, end)}{
		$features \leftarrow$ \textsc{ExtractFeatures($class$)}\;
		$seq$.append($features$)\;
		$\overline{P_{\mathcal{B}}} \leftarrow mean(P(seq~|~\mathcal{B}))$\;
		$\overline{P_{\mathcal{M}}} \leftarrow mean(P(seq~|~\mathcal{M}))$\;
		$disjoint \leftarrow$ (($\overline{P_{\mathcal{B}}} \pm t\sigma$) $\cap$ ($\overline{P_{\mathcal{M}}} \pm t\sigma$) = $\emptyset$)\;
		\uIf{$end$ \textbf{and} $disjoint$}{
			\Return $\overline{P_{\mathcal{M}}} > \overline{P_{\mathcal{B}}}$ ? \texttt{rejected} : \texttt{accepted}\;
		}
		\uElseIf{$end$ \textbf{and} $\neg disjoint$}{
			\Return \texttt{rejected}\;
		}
		\uElseIf{$disjoint$ \textbf{and} $|seq| \geq l$ \textbf{and} $\overline{P_{\mathcal{M}}} > \overline{P_{\mathcal{B}}}$}{
			\Return \texttt{rejected}\;
		}
		\Else{
			\Return \texttt{undecided}\;
		}
	}
	\caption{Deserialization Attack Prevention}
	\label{alg:predict}
\end{algorithm}

\begin{figure}
	\includegraphics[width=0.95\columnwidth]{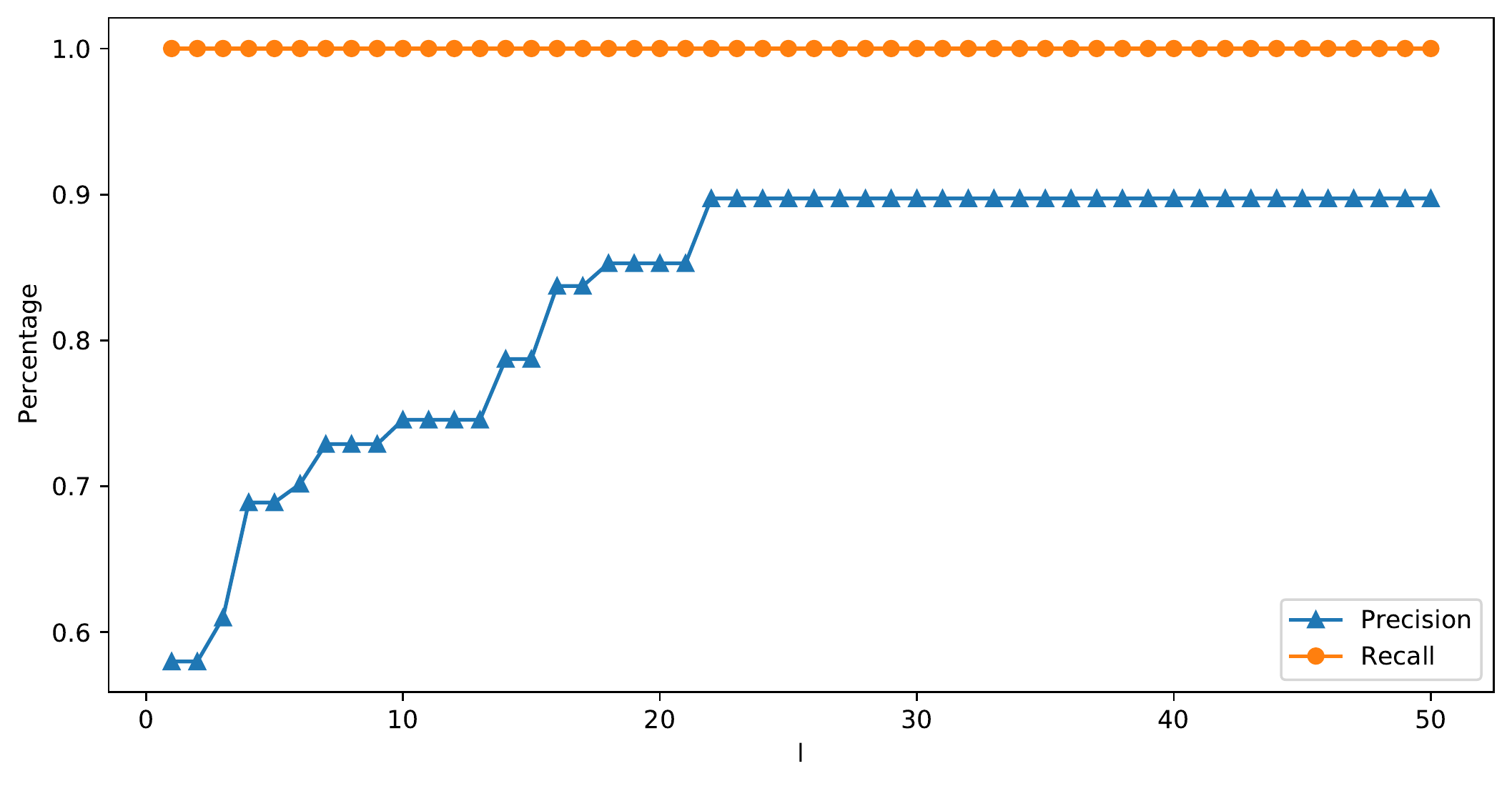}
	\caption{Average precision and recall in function of $l$}
	\label{fig:prec_rec}
\end{figure}

To validate our approach, we conducted an experiment on the Oracle 
WebLogic Server\footnote{\url{https://www.oracle.com/au/middleware/technologies/fusionmiddleware-downloads.html}}
\footnote{Oracle\textregistered WebLogic Server is a registered trademark of Oracle and/or its affiliates. Other names may be trademarks of their respective owners.} 
(WLS). 
In our setup, we collected 227 benign deserialization chains (avg. length of 38.96) from trusted runs of WLS. We 
also collected 37 malicious chains (avg. length of 16.68) from the deserialization payloads available in~\cite{ysoserial.git}. 
To measure the precision,
recall and F1-score of our approach, we conduct a 5-fold cross-validation experiment where 80\% of the
examples are used for inference and the remaining 20\% are used for prediction. Furthermore, to assess the
benefits of using statistical inference to estimate probabilities, we also conduct the same experiment 
using Markov chains inferred directly from empirical data. We use \textsc{PyMC3} for Bayesian 
inference~\cite{salvatier2016probabilistic} and \textsc{Pomegranate} for empirical 
inference~\cite{schreiber2017pomegranate}. Table~\ref{tab:results} shows the results of our experiment
with $t \in [0,3]$ and $l = \infty$, drawing 5\,000 samples from a Metropolis-Hastings sampler and using 
the last 500 samples for prediction. Bayesian inference performs significantly better, at the expense 
of inference times that are orders of magnitude larger. Note, however, that inference can be performed 
offline and that the actual runtime overhead, in the order of milliseconds, is similar for both approaches 
(i.e. extracting class features and resolving \autoref{eq:probability}).
In \autoref{alg:predict}, we let the filter stop the deserialization of malicious streams early if at 
least $l$ classes have been read and the end of the stream has not been reached. Figure \ref{fig:prec_rec} 
shows the precision and recall achieved with different values of $l$, and $t = 2$. Reading more classes is 
beneficial to precision while recall remains largely unaffected. To our knowledge, this 
is the first study to use class features and ordering for \emph{defensive} purposes against deserialization 
attacks. Considering our previous failed experiment with order-agnostic classifiers and our current F1-score
of 0.94 using MC, our results suggest that class features and ordering do capture the \emph{esssence} of malicious chains.

\begin{table}
	\caption{Precision, recall, F1-score, and inference time of Bayesian and empirical Markov chains}
	\centering
	\begin{tabular}{lcrrrr}
		\toprule
		& \multirow{2}{*}{$t$} & \multirow{2}{*}{Precision} & \multirow{2}{*}{Recall} & \multirow{2}{*}{F1-score} & Time\\
		& & & & & (sec) \\
		\midrule
		\multirow{4}{*}{Bayesian} & 0 & 91.67$\pm$6.97 & 96.67$\pm$6.67 & 0.94$\pm$0.03 &  \multirow{4}{*}{7163}\\
		& 1 & 91.67$\pm$6.97 & 96.67$\pm$6.67 & 0.94$\pm$0.03 & \\
		& 2 & 89.72$\pm$8.94 & 100.0$\pm$0.00 & 0.94$\pm$0.05 & \\
		& 3 & 88.17$\pm$11.26 & 100.0$\pm$0.00 & 0.93$\pm$0.07 &\\
		\midrule
		Empirical & --- & 72.95$\pm$14.27 & 100.0$\pm$0.00 & 0.84$\pm$0.09 & 0.7\\
		\bottomrule
	\end{tabular}
	\label{tab:results}
\end{table}

\section{Related Work}

In the security community, the classes used in a deserialization payload are referred to as 
"gadgets" and the resulting byte stream is known as a "gadget chain". Many existing work tackled
the problem of detecting gadgets chains in application and libraries
~\cite{shcherbakov2021serialdetector,dahse2014code,peles2015one,rasheed2020hybrid,haken2018automated} 
using techniques like debugger-assisted manual analysis, static analysis, and hybrid (i.e. static and
dynamic) analysis.

More closely related to our work are approaches aimed at detecting and preventing deserialization
attacks~\cite{cristalli2018trusted,pan2019detecting}. Cristalli et al. \cite{cristalli2018trusted},
present a two-phase approach that learns trusted execution paths and sandboxes the native
Java deserialization mechanism to allow deserialization from those paths only. This approach does 
not generalise to previously unseen paths and the precision thus depends on the exhaustiveness of 
the training phase. To implement their system, the authors modified a Java Virtual Machine (JVM),
and report overheads in the order of 20\%-40\%. Pan et al. \cite{pan2019detecting},
use heavyweight instrumentation (e.g. 100x slowdown) to dynamically collect execution traces 
and train deep learning models to detect malicious deserialization at runtime. To achieve an F1-score 
$>$ 0.90, authors had to manually generate over 8\,000 execution traces, which is highly unpractical. 
In contrast, our approach uses a native JVM and Java deserialization filters, and incurs a very small overhead. 

\section{Future Plans}

The work presented in this paper is in the very early stages and warrants several caveats. Our 
evaluation is currently limited to one application (WLS), one technology (Java deserialization) 
and one sampler (Metropolis-Hastings) only. Other applications, samplers, and languages will have to 
be investigated. Despite these limitations, however, we have uncovered several avenues 
for further investigation. First, our results suggest that class features and ordering capture the
essence of a malicious gadget chain. While attackers can obviously manipulate the stream to evade 
detection, successful exploitation \emph{requires} specific features and ordering. We believe that 
this invariant could be key in achieving robustness in the face of adversarial attacks. Second, 
while our approach seems promising on small datasets, it remains unclear how well it scales up and 
down and how well it applies in various user scenarios. An empirical evaluation using different workloads, 
applications and user scenarios is needed to assess the practicality of our approach.

%
%


\bibliographystyle{ACM-Reference-Format}
\bibliography{refs}


\begin{thebibliography}{18}


\ifx \showCODEN    \undefined \def \showCODEN     #1{\unskip}     \fi
\ifx \showDOI      \undefined \def \showDOI       #1{#1}\fi
\ifx \showISBNx    \undefined \def \showISBNx     #1{\unskip}     \fi
\ifx \showISBNxiii \undefined \def \showISBNxiii  #1{\unskip}     \fi
\ifx \showISSN     \undefined \def \showISSN      #1{\unskip}     \fi
\ifx \showLCCN     \undefined \def \showLCCN      #1{\unskip}     \fi
\ifx \shownote     \undefined \def \shownote      #1{#1}          \fi
\ifx \showarticletitle \undefined \def \showarticletitle #1{#1}   \fi
\ifx \showURL      \undefined \def \showURL       {\relax}        \fi
\providecommand\bibfield[2]{#2}
\providecommand\bibinfo[2]{#2}
\providecommand\natexlab[1]{#1}
\providecommand\showeprint[2][]{arXiv:#2}

\bibitem[\protect\citeauthoryear{??}{jep}{2021}]%
        {jep290}
 \bibinfo{year}{2021}\natexlab{}.
\newblock \bibinfo{title}{{JEP 290: Filter Incoming Serialization Data}}.
\newblock \bibinfo{howpublished}{\url{https://openjdk.java.net/jeps/290}}.
\newblock


\bibitem[\protect\citeauthoryear{??}{owa}{2022}]%
        {owasp.top10}
 \bibinfo{year}{2022}\natexlab{}.
\newblock \bibinfo{title}{Top 10 Web Application Security Risks}.
\newblock \bibinfo{howpublished}{\url{https://owasp.org/www-project-top-ten/}}.
\newblock


\bibitem[\protect\citeauthoryear{Bechler}{Bechler}{2017}]%
        {marshall}
\bibfield{author}{\bibinfo{person}{Moritz Bechler}.}
  \bibinfo{year}{2017}\natexlab{}.
\newblock \bibinfo{title}{{Java Unmarshaller Security: Turning your data into
  code execution}}.
\newblock
\newblock
\newblock
\shownote{\url{https://www.github.com/mbechler/marshalsec/blob/master/marshalsec.pdf}}.


\bibitem[\protect\citeauthoryear{Bruneton, Lenglet, and Coupaye}{Bruneton
  et~al\mbox{.}}{2002}]%
        {bruneton2002asm}
\bibfield{author}{\bibinfo{person}{Eric Bruneton}, \bibinfo{person}{Romain
  Lenglet}, {and} \bibinfo{person}{Thierry Coupaye}.}
  \bibinfo{year}{2002}\natexlab{}.
\newblock \showarticletitle{ASM: a code manipulation tool to implement
  adaptable systems}.
\newblock \bibinfo{journal}{\emph{Adaptable and extensible component systems}}
  \bibinfo{volume}{30}, \bibinfo{number}{19} (\bibinfo{year}{2002}).
\newblock


\bibitem[\protect\citeauthoryear{Cristalli, Vignati, Bruschi, and
  Lanzi}{Cristalli et~al\mbox{.}}{2018}]%
        {cristalli2018trusted}
\bibfield{author}{\bibinfo{person}{Stefano Cristalli}, \bibinfo{person}{Edoardo
  Vignati}, \bibinfo{person}{Danilo Bruschi}, {and} \bibinfo{person}{Andrea
  Lanzi}.} \bibinfo{year}{2018}\natexlab{}.
\newblock \showarticletitle{Trusted execution path for protecting java
  applications against deserialization of untrusted data}. In
  \bibinfo{booktitle}{\emph{International Symposium on Research in Attacks,
  Intrusions, and Defenses}}. Springer, \bibinfo{pages}{445--464}.
\newblock


\bibitem[\protect\citeauthoryear{Dahse, Krein, and Holz}{Dahse
  et~al\mbox{.}}{2014}]%
        {dahse2014code}
\bibfield{author}{\bibinfo{person}{Johannes Dahse}, \bibinfo{person}{Nikolai
  Krein}, {and} \bibinfo{person}{Thorsten Holz}.}
  \bibinfo{year}{2014}\natexlab{}.
\newblock \showarticletitle{Code reuse attacks in php: Automated pop chain
  generation}. In \bibinfo{booktitle}{\emph{Proceedings of the 2014 ACM SIGSAC
  Conference on Computer and Communications Security}}.
  \bibinfo{pages}{42--53}.
\newblock


\bibitem[\protect\citeauthoryear{Frohoff}{Frohoff}{2015}]%
        {ysoserial.git}
\bibfield{author}{\bibinfo{person}{Chris Frohoff}.}
  \bibinfo{year}{2015}\natexlab{}.
\newblock \bibinfo{title}{{ysoserial}}.
\newblock \bibinfo{howpublished}{\url{https://github.com/frohoff/ysoserial}}.
\newblock


\bibitem[\protect\citeauthoryear{Frohoff}{Frohoff}{2016}]%
        {ysoserial2}
\bibfield{author}{\bibinfo{person}{Chris Frohoff}.}
  \bibinfo{year}{2016}\natexlab{}.
\newblock \bibinfo{title}{{Deserialize My Shorts: Or How I Learned To Start
  Worrying and Hate Java Object Deserialization}}.
\newblock
\newblock
\newblock
\shownote{\url{http://frohoff.github.io/owaspsd-deserialize-my-shorts/}}.


\bibitem[\protect\citeauthoryear{Goetz}{Goetz}{2019}]%
        {TBS}
\bibfield{author}{\bibinfo{person}{Brian Goetz}.}
  \bibinfo{year}{2019}\natexlab{}.
\newblock \bibinfo{title}{{Towards Better Serialization}}.
\newblock
\newblock
\newblock
\shownote{\url{https://cr.openjdk.java.net/~briangoetz/amber/serialization.html}}.


\bibitem[\protect\citeauthoryear{Haken}{Haken}{2018}]%
        {haken2018automated}
\bibfield{author}{\bibinfo{person}{Ian Haken}.}
  \bibinfo{year}{2018}\natexlab{}.
\newblock \bibinfo{title}{Automated Discovery of Deserialization Gadget
  Chains}.
\newblock
\newblock


\bibitem[\protect\citeauthoryear{Lawrence and Frohoff}{Lawrence and
  Frohoff}{2015}]%
        {ysoserial}
\bibfield{author}{\bibinfo{person}{Gabriel Lawrence} {and}
  \bibinfo{person}{Chris Frohoff}.} \bibinfo{year}{2015}\natexlab{}.
\newblock \bibinfo{title}{{Marshalling Pickles: How Deserializing Objects Can
  Ruin Your Day}}.
\newblock
\newblock
\newblock
\shownote{\url{https://frohoff.github.io/appseccali-marshalling-pickles}}.


\bibitem[\protect\citeauthoryear{Mu{\~{n}}oz and Mirosh}{Mu{\~{n}}oz and
  Mirosh}{2017}]%
        {Deser.NET}
\bibfield{author}{\bibinfo{person}{Alvaro Mu{\~{n}}oz} {and}
  \bibinfo{person}{Oleksandr Mirosh}.} \bibinfo{year}{2017}\natexlab{}.
\newblock \bibinfo{title}{{Friday the 13th JSON Attacks}}.
\newblock
\newblock
\newblock
\shownote{\url{https://www.blackhat.com/docs/us-17/thursday/us-17-Munoz-Friday-The-13th-JSON-Attacks-wp.pdf}}.


\bibitem[\protect\citeauthoryear{Pan, Sun, Teng, White, Schmidt, Staples, and
  Krause}{Pan et~al\mbox{.}}{2019}]%
        {pan2019detecting}
\bibfield{author}{\bibinfo{person}{Yao Pan}, \bibinfo{person}{Fangzhou Sun},
  \bibinfo{person}{Zhongwei Teng}, \bibinfo{person}{Jules White},
  \bibinfo{person}{Douglas~C Schmidt}, \bibinfo{person}{Jacob Staples}, {and}
  \bibinfo{person}{Lee Krause}.} \bibinfo{year}{2019}\natexlab{}.
\newblock \showarticletitle{Detecting web attacks with end-to-end deep
  learning}.
\newblock \bibinfo{journal}{\emph{Journal of Internet Services and
  Applications}} \bibinfo{volume}{10}, \bibinfo{number}{1}
  (\bibinfo{year}{2019}), \bibinfo{pages}{1--22}.
\newblock


\bibitem[\protect\citeauthoryear{Peles and Hay}{Peles and Hay}{2015}]%
        {peles2015one}
\bibfield{author}{\bibinfo{person}{Or Peles} {and} \bibinfo{person}{Roee Hay}.}
  \bibinfo{year}{2015}\natexlab{}.
\newblock \showarticletitle{One class to rule them all: 0-day deserialization
  vulnerabilities in android}. In \bibinfo{booktitle}{\emph{9th $\{$USENIX$\}$
  Workshop on Offensive Technologies ($\{$WOOT$\}$ 15)}}.
\newblock


\bibitem[\protect\citeauthoryear{Rasheed and Dietrich}{Rasheed and
  Dietrich}{2020}]%
        {rasheed2020hybrid}
\bibfield{author}{\bibinfo{person}{Shawn Rasheed} {and} \bibinfo{person}{Jens
  Dietrich}.} \bibinfo{year}{2020}\natexlab{}.
\newblock \showarticletitle{A hybrid analysis to detect Java serialisation
  vulnerabilities}. In \bibinfo{booktitle}{\emph{Proceedings of the 35th
  IEEE/ACM International Conference on Automated Software Engineering}}.
  \bibinfo{pages}{1209--1213}.
\newblock


\bibitem[\protect\citeauthoryear{Salvatier, Wiecki, and Fonnesbeck}{Salvatier
  et~al\mbox{.}}{2016}]%
        {salvatier2016probabilistic}
\bibfield{author}{\bibinfo{person}{John Salvatier}, \bibinfo{person}{Thomas~V
  Wiecki}, {and} \bibinfo{person}{Christopher Fonnesbeck}.}
  \bibinfo{year}{2016}\natexlab{}.
\newblock \showarticletitle{{Probabilistic Programming in Python using PyMC3}}.
\newblock \bibinfo{journal}{\emph{PeerJ Computer Science}}  \bibinfo{volume}{2}
  (\bibinfo{year}{2016}), \bibinfo{pages}{e55}.
\newblock


\bibitem[\protect\citeauthoryear{Schreiber}{Schreiber}{2017}]%
        {schreiber2017pomegranate}
\bibfield{author}{\bibinfo{person}{Jacob Schreiber}.}
  \bibinfo{year}{2017}\natexlab{}.
\newblock \showarticletitle{{Pomegranate: Fast and Flexible Probabilistic
  Modeling in Python}}.
\newblock \bibinfo{journal}{\emph{The Journal of Machine Learning Research}}
  \bibinfo{volume}{18}, \bibinfo{number}{1} (\bibinfo{year}{2017}),
  \bibinfo{pages}{5992--5997}.
\newblock


\bibitem[\protect\citeauthoryear{Shcherbakov and Balliu}{Shcherbakov and
  Balliu}{2021}]%
        {shcherbakov2021serialdetector}
\bibfield{author}{\bibinfo{person}{Mikhail Shcherbakov} {and}
  \bibinfo{person}{Musard Balliu}.} \bibinfo{year}{2021}\natexlab{}.
\newblock \showarticletitle{SerialDetector: Principled and Practical
  Exploration of Object Injection Vulnerabilities for the Web}. In
  \bibinfo{booktitle}{\emph{Network and Distributed Systems Security (NDSS)
  Symposium 202121-24 February 2021}}.
\newblock


\end{thebibliography}

\end{document}